\documentstyle[aps,prl,multicol,ifthen,epsf]{revtex}
\tighten

\begin{document}

\draft

\title{Geometrically constrained magnetic wall}

\author{P. Bruno\cite{e-mail}}

\address{Max-Planck-Institut f\"ur Mikrostrukturphysik, 
Weinberg 2, D-06120 Halle, Germany}

\date{26 April 1999; revised 19 July 1999}

\maketitle

\begin{abstract}
The structure and properties of a geometrically constrained magnetic wall in 
a constriction separating two wider regions are studied theoretically. 
They are shown to differ considerably from those of an 
unconstrained wall, so that the geometrically 
constrained magnetic wall truly constitutes 
a new kind of magnetic wall, besides the well known Bloch and N\'eel walls. 
In particular, the width of a constrained wall 
can become very small if 
the characteristic length of the constriction is small, as is actually the 
case in an atomic point contact. This provides a simple, natural explanation 
for the large magnetoresistance observed recently in 
ferromagnetic atomic point contacts.
\end{abstract}

\pacs{Published in: Phys. Rev. Lett. {\bf 83}, 2425--2428 (1999)}

\begin{multicols}{2}

The investigation of magnetic nanostructures is one of the major current 
subjects in magnetism. This interest is stimulated on one hand, by the 
great progress in nanofabrication techniques and magnetic characterization 
methods, and on the other hand by the perspective of technological 
applications for magnetic storage of information of unprecedented density.

A major question to be addressed in this field of research is: how does the 
micromagnetic structure (i.e., domains, walls, etc.) respond to 
geometrical constraints on the nanometer scale? 

In an unconstrained system such as in a bulk ferromagnet, as first pointed out 
by F.~Bloch, the wall structure is determined by a competition between 
exchange and anisotropy energies \cite{ Bloch1932}. The exact structure of the 
Bloch wall has been calculated  
by Landau and Lifshitz \cite{ Landau1935}. In a ferromagnetic thin film 
with in-plane easy magnetization axis, as shown by N\'eel \cite{ Neel1955}, 
the dipolar interaction leads to new kind of wall, known as a N\'eel wall,
in which the structure is determined by a competition between exchange, 
anisotropy, and dipolar energies.

In this Letter, I consider the problem of the structure and energy of a 
magnetic wall in a constriction separating two regions of 
wider cross-section.
This encompasses various situations of great physical interest such as a 
narrow constriction fabricated in an magnetic ultrathin film by lithographic 
techniques, or a constriction in a wire.

I point out that when the cross-section of the constriction is much smaller 
than that of the wide region, the structure of the wall becomes almost 
independent of the material parameters such as magnetization, exchange 
stiffness and anisotropy constant, and is determined mostly by the geometry 
of the constriction. The wall energy consists mostly of exchange energy. 
Thus, geometrically constrained magnetic walls appear as a new kind of magnetic 
walls, with properties completely different from those of Bloch and N\'eel 
walls. In particular, the width of the geometrically constrained magnetic walls 
is essentially given by the length of the constriction, which can be 
considerably 
smaller than the width of a Bloch or N\'eel wall. In the limit of a point 
contact of atomic dimensions, the width of the wall would also be of atomic 
dimensions. This fact has important physical consequences: in particular, the 
contribution of a geometrically constrained magnetic wall to the electrical 
resistance of a point contact will be considerably larger than that 
of a Bloch or N\'eel wall, so that very large magnetoresistance effects 
can be anticipated in ferromagnetic point contacts 
and have actually been reported 
\cite{ Garcia1999}.

Let us  consider an homogeneous magnetic system in which the cross-section 
$S(x)$ 
varies along the $x$ axis and exhibits a minimum at $x=0$. The easy 
magnetization axis is along the $z$ axis and the 
magnetization for $x \to \pm\infty$ is along the $\pm z$~axis, respectively. 
Obviously, the magnetic wall will tend to localize itself 
near the constriction, in 
order to minimize its energy. In order to understand how the the wall structure 
is modified by the constriction, we can make the following reasoning: let us 
start with an infinitely narrow wall located at the center of the constriction. 
The exchange energy of such a configuration is too high and can be reduced 
by allowing the wall to expand. This expansion is counterbalanced by (i)
the increase of anisotropy and (ii) by the increase of wall area. If 
the cross-section $S(x)$ increases rapidly with $|x|$, then the second term 
can be the leading one, so that the wall structure will be controlled 
essentially by the geometry of the constriction, and depend only weakly on the 
material parameters.

For explicit calculations below, I shall consider the following models of 
constrictions:
\begin{eqnarray}
 & &
\begin{array}{llllr} 
\begin{array}{l}
S(x) \\
 
\end{array} &
\begin{array}{l}
= \\
=
\end{array} &
\begin{array}{l}
S_0 \\
S_1 > S_0
\end{array} &
\left. 
\begin{array}{l}
\mbox{for $|x| \le d$} \\
\mbox{for $|x| \ge d$}
\end{array} \ \ \right\} &  \mbox{(model I)} \\
S(x) &=& S_0 \left( 1+ \frac{x^2}{d^2} \right) &  &  
\mbox{(model II)} \\
S(x) &=& S_0  \, \cosh (x/d) & &   \mbox{(model III)}
\end{array} \nonumber
\end{eqnarray}

For the sake of simplification, I shall make the 
following assumptions: (i) the magnetization direction depends only on $x$, 
i.e., the wall is plane (in reality this assumption is not strictly satisfied 
and the wall would tend to bend), (ii) the dipolar interactions can be 
neglected (the validity of this assumption be discussed later), and (iii) the 
magnetization remains in the $yz$--plane like in a Bloch wall (this assumption 
is best suited to the case of a constriction in a film with perpendicular 
anisotropy; in general, however, the wall structure would deviate from the 
idealized Bloch-like configuration). 
One can argue, however, 
that the above simplications would not modify significantly the underlying 
physical mechanism, and provides a good approximation of the wall structure
and energy.  
Thus, the wall is described by the angle $\theta (x)$ between 
the magnetization and the $z$~axis. Let $F(\theta )$ be the anisotropy energy 
density and $A$ the exchange stiffness. 
In practice, we shall assume a uniaxial anisotropy below, i.e.,
$F(\theta ) \equiv K \cos^2\theta$; however, wherever we use the more general 
form $F(\theta )$, the results are not restricted to this particular case. 
The total energy of the wall is given 
by
\begin{equation}\label{eq:energy_tot}
E[\theta ] = \int_{-\infty}^{\infty} {\rm d}x \left[ A {\dot{\theta}}^2 +
F(\theta ) \right] S(x) ,
\end{equation}
where $\dot{\theta} \equiv {\rm d}\theta /{\rm d}x$. 
The structure of the 
wall is obtained by solving the corresponding Euler equation
\begin{equation}\label{eq:Euler}
\ddot{\theta} +\dot{\theta} \, \frac{\dot{S}}{S} - \frac{F^\prime (\theta )}{2A}
=0 ,
\end{equation}
where $F^\prime (\theta ) \equiv {\rm d}F/{\rm d}\theta$, subject to the 
boundary conditions, $\theta (\pm\infty ) = \pm \pi /2$, respectively, and 
$\dot{\theta} = 0$ for $\theta = \pm \pi /2$. 
The new term $\dot{\theta}\dot{S}/S$, which is absent in the case of 
an unconstrained Bloch wall 
considered by Landau and Lifshitz \cite{ Landau1935}, 
expresses the influence of 
the geometry of the constriction on the wall structure.

We shall be interested, in 
particular, in the width and energy of the wall. 
Various definitions of the wall 
width have been proposed in the literature \cite{ Hubert1998}. Here, since 
we have in mind the electrical transport properties of the wall, we need 
an appropriate definition of the wall width. As the electrical resistance of 
a magnetic wall is determined by $\dot{\theta}(x)$ \cite{Tatara1997}, 
this naturally leads 
one to use for the wall width the following new definition:
\begin{equation}\label{eq:width_def}
w \equiv 4 \left[ \int_{-\infty}^\infty {\dot{\theta}}^2(x) \, {\rm d}x 
\right]^{-1}  = 4 \left[ \int_{-\pi /2}^{\pi /2} \dot{\theta}\, {\rm d}\theta 
\right]^{-1} ,
\end{equation}
where the prefactor has been chosen so that this definition yields 
$w_0= 2 \sqrt{A/K}$ for the unconstrained Bloch wall.

Let us neglect the term $-F^\prime (\theta)/A$ in eq.~(\ref{eq:Euler}), and 
call $\theta^\star (x)$ the corresponding solution, of width $w^\star$ 
and energy $E^\star$.
This provides a good approximation of the true solution if $(\dot{S}/S)^2$ is 
large as compared to $|F^\prime (\theta )|/A$, and in any case yields an upper 
limit for the wall width. The solution then take the general form 
\begin{equation}\label{eq:theta_star}
\theta^\star(x) = \pi \left[ \frac{\int_{-\infty}^x S^{-1}(x^\prime ) 
{\rm d}x^\prime}{ \int_{-\infty}^\infty S^{-1}(x^\prime ) 
{\rm d}x^\prime} - \frac{1}{2} \right] .
\end{equation}
The width is given by
\begin{equation}\label{eq:w_star}
w^\star =\frac{4}{\pi^2} \frac{\left[ \int_{-\infty}^{\infty} S^{-1}(x) \, 
{\rm d}x
\right]^2}{\int_{-\infty}^{\infty} S^{-2}(x) \, {\rm d}x} 
\end{equation}
and the wall energy is
\begin{equation}\label{eq:e_star}
E^\star = \frac{\pi^2\, A}{\int_{-\infty}^\infty S^{-1}(x)\, {\rm d}x} .
\end{equation}

In fact, we can argue that the above approximation is justified whenever 
$w^\star$ is small as compared to the width $w_0 = 2 \sqrt{A/K}$ 
of the unconstrained Bloch 
wall, which is the only relevant parameter characterizing the material. 
Obviously, the usefulness of this approximation depends on whether the 
integral $\int_{-\infty}^{+\infty} S^{-1}(x)\, {\rm d}x$ converges or not 
(for brievety, we shall term the constriction, respectively ``integrable'' 
and ``non integrable'').
If $S(x)$ diverges more rapidly than 
$|x|$ for $|x| \to \pm \infty$, the constriction is ``integrable''; 
this is the case for model I with 
$S_1=\infty$, as well as for models II 
and III. In the opposit case of a ``non integrable'' constriction 
(e.g., model I with $S_1$ finite), $w^\star = \infty$, so that 
the term $-F^\prime (\theta)/A$ can never be neglected. 

Let us first discuss the case of an ``integrable'' constriction 
with $w^\star \ll w_0$. In this case, 
the above discussion shows that the wall structure is {\em independent\/} 
of the material parameter $w_0$ and is determined only by the geometry of 
constriction; furthermore, the 
wall profile $\theta^\star(x)$ and the wall width $w^\star$ are 
independent of the constriction cross-section $S_0$. 
The area of the constriction is irrelevant; the only relevant parameter 
is the length $d$ on which $S(x)$ varies significantly. Thus, we expect 
that the width will be of the order of $w^\star \sim d$ and the 
wall energy of the order of $AS_0/d$. Performing explicitely 
the calculation, this yields, for model I with $S_1=\infty$,
\begin{mathletters}
\begin{eqnarray}
\theta^\star (x) &=& \frac{\pi x}{2d} , \\
w^\star &=& \frac{8d}{\pi^2} , \label{eq:model1}\\
E^\star &=& \frac{\pi^2 A S_0}{2d} ,
\end{eqnarray}
\end{mathletters}
whereas for model II one gets 
\begin{mathletters}
\begin{eqnarray}
\theta^\star(x) &=& \arctan (x/d) , \\
w^\star &=& \frac{8d}{\pi} ,  \\
E^\star &=& \frac{\pi A S_0}{d} ,
\end{eqnarray}
\end{mathletters}
and for model III
\begin{mathletters}
\begin{eqnarray}
\theta^\star(x) &=& \arcsin \left[ \tanh (x/d) \right] , \\
w^\star &=& 2d \\
E^\star &=& \frac{\pi A S_0}{d} ,
\end{eqnarray}
\end{mathletters}
which confirms the above qualitative discussion.
Interestingly, we remark that, for model III, the wall profile has 
the same form as for the unconstrained Bloch wall, with $d$ replacing $L\equiv 
\sqrt{A/K}$.

\narrowtext{
\begin{figure}[t,h]
\begin{center}
\epsfxsize=8cm
\epsffile{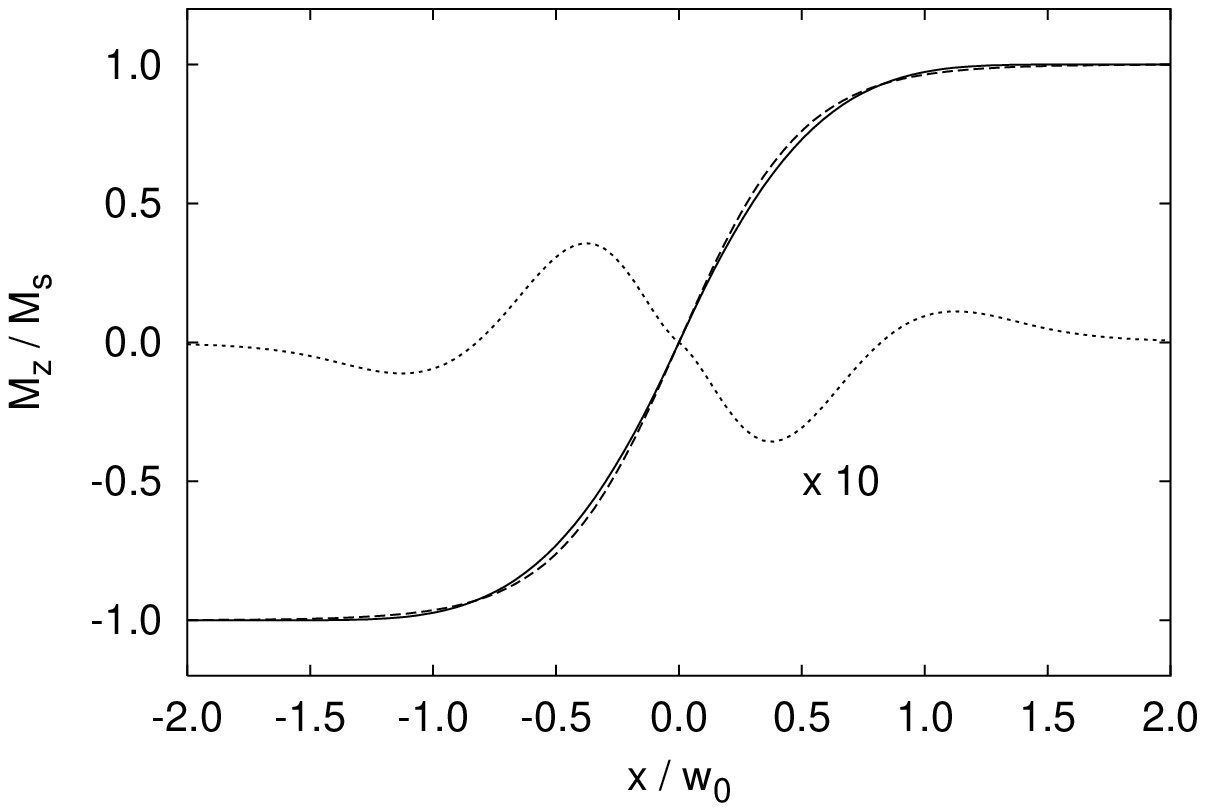}
\end{center}
\vspace*{-\baselineskip}
\caption{Magnetization profile of the unconstrained Bloch wall 
(long-dashed line), as 
compared with the one calculated using aproximation (\protect\ref{eq:approx},b)
(solid line); the difference between the approximate and exact 
solutions (magnified by a factor 10) is also show (short-dashed line).}
\label{fig:Bloch}
%\end{figure}
%
%\begin{figure}[h]
\begin{center}
\epsfxsize=8cm
\epsffile{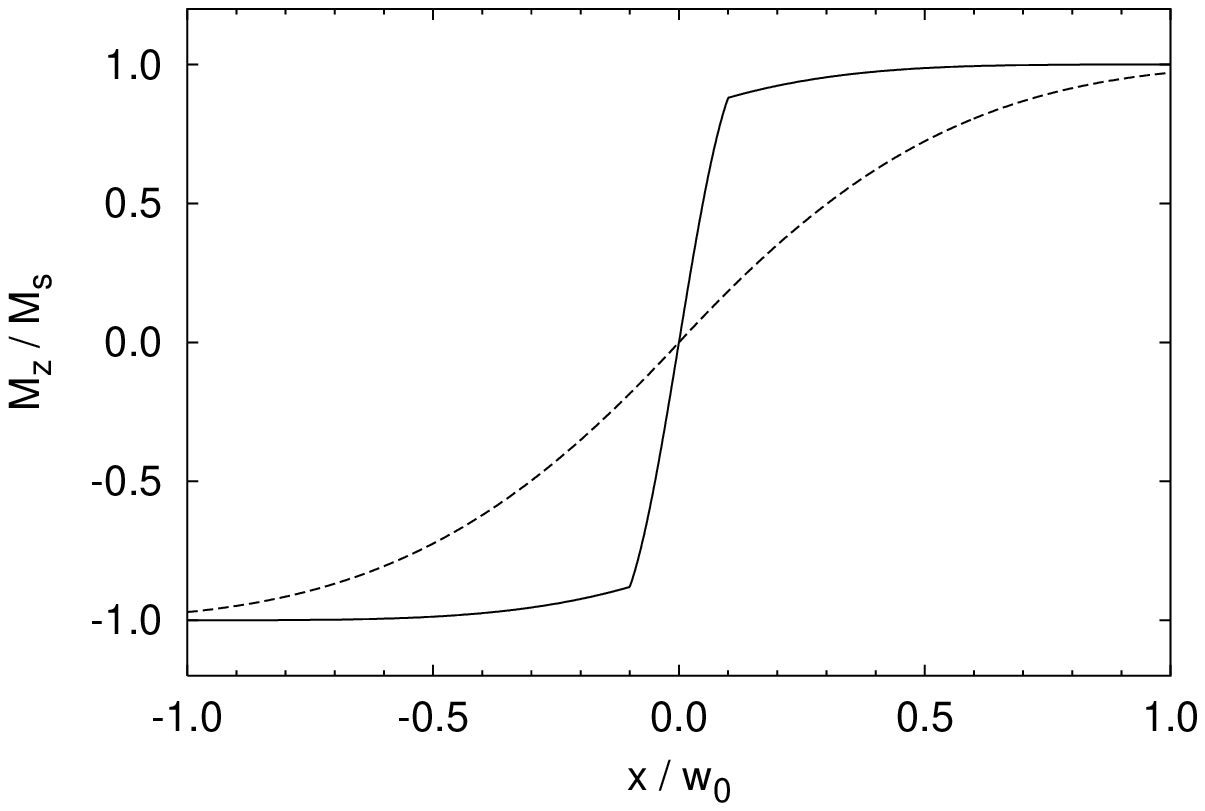}
\end{center}
\vspace*{-\baselineskip}
\caption{Magnetization profile of a geometrically constrained magnetic 
wall calculated for model I with $d/w_0 =0.1$ 
and $S_1/S_0 = 10$ (solid line), as compared with to the unconstrained 
Bloch wall 
(dashed 
line).}
\label{fig:constrained}
\end{figure}
}

On the other hand, if the constriction is ``non integrable'', or if $w^\star$ 
is not small as compared to $w_0$, the term 
$-F^\prime(\theta )/A$ cannot be neglected {\em a priori\/}; 
thus, this case deserves a 
more careful study. We now specify to the case of uniaxial anisotropy. 
The Euler equation becomes
\begin{equation}\label{eq:Euler2}
\ddot{\theta} +\dot{\theta} \, \frac{\dot{S}}{S} + \frac{\sin\theta \cos\theta}
{L^2}=0 .
\end{equation}
In order to make this equation easily soluble, we perform the 
following approximation
\begin{mathletters}
\begin{eqnarray}\label{eq:approx}
\cos^2\theta &\approx& \alpha\, (\pi - 2 |\theta |) \\
\sin\theta \cos\theta &\approx& \alpha\ \mbox{sgn}(\theta) ,
\end{eqnarray}
\end{mathletters}
for $|\theta| \le \pi/2$, and 
where the parameter $\alpha$ is determined variationally by minimizing the 
energy with respect to $\alpha$ for the unconstrained Bloch wall, which yields 
$\alpha = 0.298901\dots$. In spite of its simplicity and its 
apparent crudeness, this approximation 
is an excellent one and yields a wall profile which is almost identical 
to exact one, as can be seen in Fig.~\ref{fig:Bloch}, 
while the errors on the wall width and energy are smaller than 
1.5~\%.

With the help of this approximation, it is straightforward (although tedious) 
to solve the Euler equation almost completely analytically. Since the resulting 
expressions are rather cumbersome \cite{ note}, 
I shall give below only approximate 
expressions valid in a restricted range of parameters, from which the 
physical meaning appears more clearly; the figures, however, display results 
obtained from the full expressions. 

The wall profile calculated for model I is shown in 
Fig.~\ref{fig:constrained}. The wall consists of a core region of width $2d$ 
in which most of the magnetization rotation takes place, and tails of width 
of the order of $w_0$ in which the magnetization rotates only weakly. Thus, if 
$d \ll w_0$, the constrained wall is much narrower than an unconstrained 
Bloch wall.

\begin{figure}[h]
\begin{center}
\epsfxsize=8cm
\epsffile{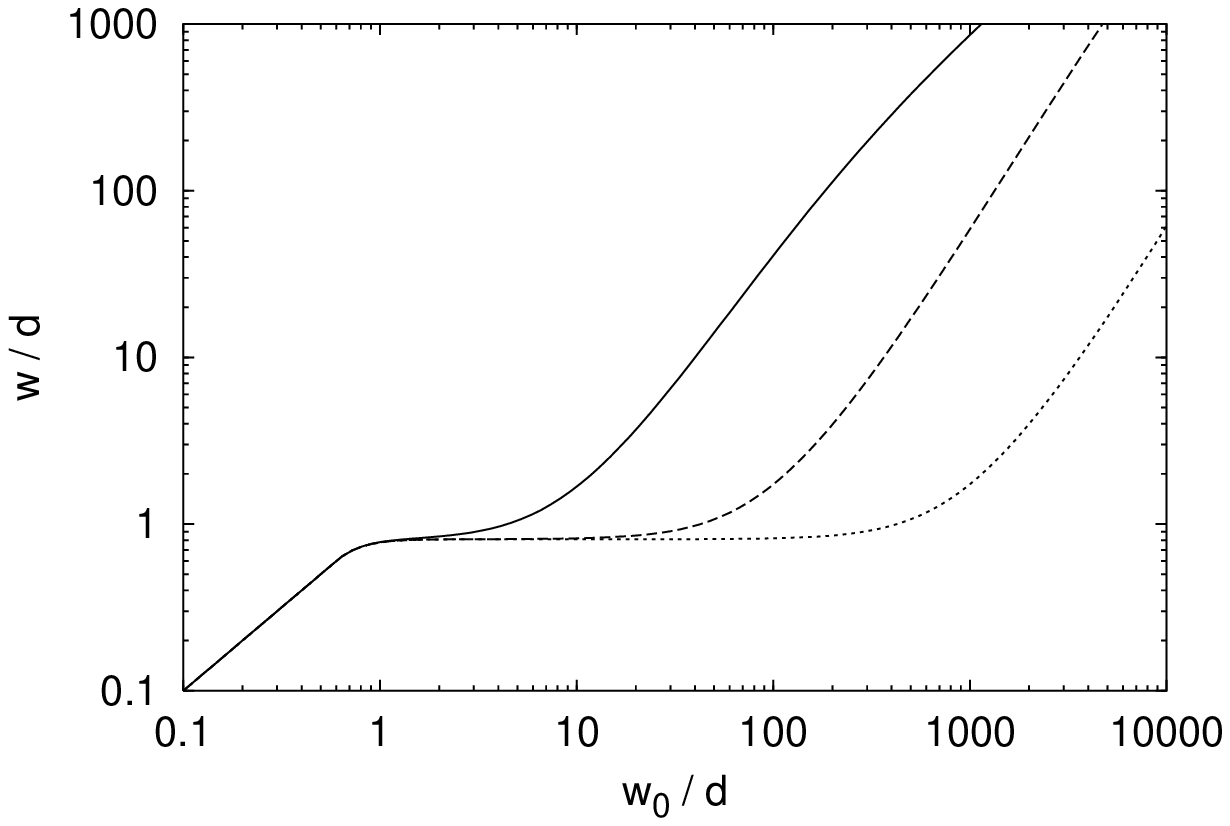}
\end{center}
\vspace*{-\baselineskip}
\caption{Wall width (normalized to $d$) of a geometrically constrained 
magnetic wall calculated for model I, as a function of 
$w_0/d$; solid line: $S_1/S_0=10$; long-dashed line: $S_1/S_0=10^2$;
short-dashed line: $S_1/S_0=10^3$.}
\label{fig:width}
%\end{figure}
%
%\begin{figure}[h]
\begin{center}
\epsfxsize=8cm
\epsffile{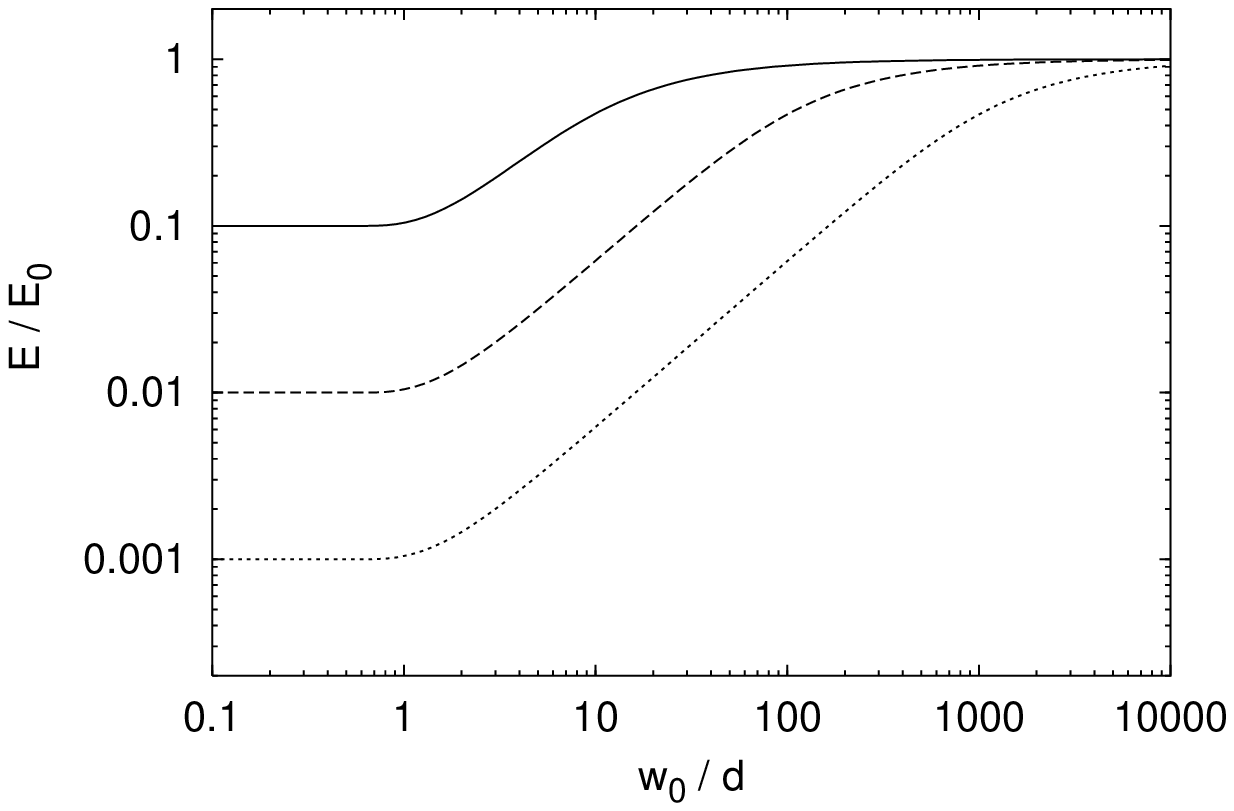}
\end{center}
\vspace*{-\baselineskip}
\caption{Wall energy (normalized to $E_0$) of a geometrically constrained 
magnetic wall calculated for model I, as a function of 
$w_0/d$; solid line: $S_1/S_0=10$; long-dashed line: $S_1/S_0=10^2$;
short-dashed line: $S_1/S_0=10^3$.}
\label{fig:energy}
\end{figure}

The wall width (normalized to $d$) and energy 
(normalized to the energy in absence of 
constriction, $E_0 = \gamma_0 \, S_1$, 
where $\gamma_0 \equiv 4 \sqrt{AK}$ is the energy per unit area of the 
unconstrained Bloch wall) as a function of $w_0 /d$ are displayed in 
Figs.~\ref{fig:width} and \ref{fig:energy}, respectively, for various values 
of the ratio $S_1/S_0$.

We can distinguish here 3 different regimes, clearly visible on 
Figs.~\ref{fig:width} and \ref{fig:energy}, depending on the values of 
the parameters $w_0/d$ and $S_1/S_0$. In the first regime, 
i.e. for $w_0/d \le 1$, one has 
\begin{mathletters}
\begin{eqnarray}
w &\approx & w_0 , \\
E &\approx & \gamma_0 S_0 .
\end{eqnarray}
\end{mathletters}
This is easily understood: since the unconstrained wall width 
$w_0$ is smaller than the 
characteristic length of the constriction, the wall is entirely confined 
in the constriction, and is therefore not significantly influenced by 
it.  

The situation is completely different in the regime characterized by 
$1 \le w_0 /d \le S_1/S_0$, for which one gets
\begin{mathletters}
\begin{eqnarray}
w &\approx& \frac{8d}{\pi^2} , \\
E &\approx& \frac{\pi^2 AS_0}{2d} ,
\end{eqnarray}
\end{mathletters}
i.e., the wall width and energy are the same as the ones obtained 
for $S_1=\infty$, (\ref{eq:model1},c), on the basis of approximations 
(\ref{eq:w_star}) and (\ref{eq:e_star}). Here the wall structure and 
wall width depend only on the geometry of the constriction, and not at all on 
the material parameters, while the wall energy is of pure exchange character. 
If the ratio $S_1/S_0$ is large, this regime is achieved in a wide range 
of values of $w_0/d$, as appears clearly from Figs.~\ref{fig:width} and 
\ref{fig:energy}. 

Finally, for $w_0/d \ge S_1/S_0$, one gets
\begin{mathletters}
\begin{eqnarray}
w &\approx& w_0 \left[ 1+ \frac{18}{\pi^2} \frac{d}{w_0}
\left( \left(\frac{S_1}{S_0}\right)^2 - \frac{S_1}{S_0} \right) 
\right]^{-1} , \\
E &\approx& \gamma_0S_1 \left[ 1- \frac{9}{\pi^2} \frac{d}{w_0} 
\frac{S_1}{S_0}
+ \frac{54}{\pi^4} \frac{d^2}{{w_0}^2} \left( \frac{S_1}{S_0} \right)^2 
\right] .
\end{eqnarray}
\end{mathletters}
This is the case in which the wall structure is again determined 
primarily by the 
competition between the exchange and anisotropy energy terms, i.e., the first 
and third terms in the Euler equation (\ref{eq:Euler}), the additional 
term $\dot{\theta}\dot{S}/S$ due to the constriction being of 
secondary importance; therefore $w$ and $E$ tend respectively towards 
$w_0$ and $E_0$ as $w_0/d$ increases. 

Let us now discuss the r\^ole of dipolar interactions, which we have 
neglected so far. A rough estimate of the dipolar contribution to the wall 
energy is given by $2 \pi M_s^2$ multiplied by the wall volume, i.e., 
$E_d \approx 2\pi M_s^2 S_0 w$. If this energy is small as compared to 
the wall energy $E$ calculated by 
neglecting the dipolar contribution, then we can expect that dipolar 
interactions will have only a small influence on the wall structure and 
on its width. For the most interesting case where $1 \le w_0/d \le S_1/S_0$, 
one finds that dipolar interactions can be neglected if 
$d \ll (\pi^2/4) \lambda$, where $\lambda \equiv A/(2\pi M_s^2)$ is the 
exchange length. For the typical value $\lambda \approx 3$~nm, this means 
$d \ll 7.5$~nm. Thus, for an atomic point contact, our approximation is well 
justified. 

In conclusion, I have investigated the properties of a geometrically 
constrained magnetic 
wall. I have shown that the structure and the properties 
of such a wall differ considerably from those of an 
unconstrained wall, so that the geometrically 
constrained magnetic wall truly constitutes 
a new kind of magnetic wall, besides the well known Bloch and N\'eel walls. 
In particular, the width of a geometrically constrained magnetic wall 
can become very small if 
the characteristic length of the constriction is small, as is actually the 
case in an atomic point contact. This provides a simple, natural explanation 
for the large magnetoresistance which has been recently observed in atomic 
point contacts \cite{ Garcia1999}. 
In addition, I have introduced a new definition of the wall 
thickness, which is the appropriate one for discussing the electrical 
resistance of magnetic wall. I have also proposed a simple approximation 
for solving the 
Euler equation, which allows to obtain simplified, yet accurate, analytical 
results.

I wish to express my gratitude to 
Gen Tatara, Hans-Peter Oepen, Yonko Millev, 
as well as to the late 
Alex Hubert, for helpful discussions.

\end{multicols}


\begin{references}
\vspace*{-1cm}

\bibitem[*]{e-mail} Electronic address: {\tt bruno@mpi-halle.de}

\bibitem{ Bloch1932} F.~Bloch, Z. Phys. {\bf 74}, 295 (1932).

\bibitem{ Landau1935} L.D.~Landau and E.M.~Lifshitz, Phys. Z. Sowietunion
{\bf 8}, 153 (1935).

\bibitem{ Neel1955} L.~N\'eel, C.R. Acad. Sc. (Paris) {\bf 241}, 533 (1955).

\bibitem{ Garcia1999} N.~Garc\'\i a, M.~Mu\~ noz, and Y.-W.~Zhao, 
Phys. Rev. Lett. {\bf 82}, 2923 (1999).

\bibitem{ Hubert1998} A.~Hubert and R.~Sch\"afer, {\em Magnetic Domains\/}
(Springer Verlag, Berlin, 1998).

\bibitem{Tatara1997} G.~Tatara and H.~Fukuyama, Phys. Rev. Lett. {\bf 78}, 
3773 (1997).

\bibitem{ note} Explicitely, one obtains $w/w_0 = (\omega^3 + 3\delta 
\omega^2\sigma^2 + 3\omega\delta^2\sigma + \delta^3)^{-1}$ and 
$E/E_0 = \omega^3+3\omega^2\delta\sigma/2 +3\delta/(2\sigma)-
\delta^3/(2\sigma)$ for $\delta \le 1$, and $w/w_0=1$ and $E / E_0 = 1/ 
\sigma$ for $\delta \ge 1$. 
Here, $\sigma \equiv S_1/S_0$, $\delta \equiv 6d/(\pi^2 w_0)$, and 
$\omega \equiv \sigma [-\delta + \sqrt{\delta^2 + (1-\delta^2)/\sigma^2}]$.

\end{references}
\end{document}